\documentclass[sigconf]{acmart}

\usepackage{censor}
\usepackage{enumitem}
\usepackage{hyperref}
\definecolor{linkColor}{RGB}{56,115,203}

\AtBeginDocument{%
  }

\copyrightyear{2025}
\acmYear{2025}
\setcopyright{rightsretained}
\acmConference[SIGCSE TS 2025]{Proceedings of the 56th ACM Technical
Symposium on Computer Science Education V. 2}{February 26-March 1,
2025}{Pittsburgh, PA, USA}
\acmBooktitle{Proceedings of the 56th ACM Technical Symposium on Computer
Science Education V. 2 (SIGCSE TS 2025), February 26-March 1, 2025,
Pittsburgh, PA, USA}
\acmDOI{10.1145/3641555.3705217}
\acmISBN{979-8-4007-0532-8/25/02}

\begin{document}

\title{Managing Project Teams in an  Online Class of 1000+ Students}

\author{Nazanin Tabatabaei Anaraki}
\email{nazanin.tbt@gatech.edu}
\affiliation{%
  \institution{Georgia Institute of Technology}
  \city{Atlanta}
  \state{Georgia}
  \country{USA}
}
\author{Taneisha Ng}
\email{taneisha.ng@gatech.edu}
\affiliation{%
  \institution{Georgia Institute of Technology}
  \city{Atlanta}
  \state{Georgia}
  \country{USA}
}
\author{Gaurav Verma}
\email{gverma@gatech.edu}
\affiliation{%
  \institution{Georgia Institute of Technology}
  \city{Atlanta}
  \state{Georgia}
  \country{USA}
}
\author{Yu Fu}
\email{fuyu@gatech.edu}
\affiliation{%
  \institution{Georgia Institute of Technology}
  \city{Atlanta}
  \state{Georgia}
  \country{USA}
}

\author{Martin O'Connell}
\email{moconnell33@gatech.edu}
\affiliation{%
  \institution{Georgia Institute of Technology}
  \city{Atlanta}
  \state{Georgia}
  \country{USA}
}

\author{Matthew Hull}
\email{matthewhull@gatech.edu}
\affiliation{%
  \institution{Georgia Institute of Technology}
  \city{Atlanta}
  \state{Georgia}
  \country{USA}
}

\author{Susanta Routray}
\email{sroutray3@gatech.edu}
\affiliation{%
  \institution{Georgia Institute of Technology}
  \city{Atlanta}
  \state{Georgia}
  \country{USA}
}

\author{Max Mahdi Roozbahani}
\email{mahdir@gatech.edu}
\affiliation{%
  \institution{Georgia Institute of Technology}
  \city{Atlanta}
  \state{Georgia}
  \country{USA}
}

\author{Duen Horng Chau}
\email{polo@gatech.edu}
\affiliation{%
  \institution{Georgia Institute of Technology}
  \city{Atlanta}
  \state{Georgia}
  \country{USA}
}

\renewcommand{\shortauthors}{Nazanin Tabatabaei Anaraki et al.}

\begin{abstract}
Team projects in Computer Science (CS) help students build collaboration skills, apply theory, and prepare for real-world software development. Online classes present unique opportunities to transform the accessibility of CS education at scale. Still, the geographical distribution of students and staff adds complexity to forming effective teams, providing consistent feedback, and facilitating peer interactions.  We discuss our approach of managing, evaluating, and providing constructive feedback to over 200 project teams, comprising 1000+ graduate students distributed globally, two professors, and 25+ teaching assistants. We deployed and iteratively refined this approach over 10 years while offering the Data and Visual Analytics course (CSE 6242) at Georgia Institute of Technology.  Our approach and insights can help others striving to make CS education accessible, especially in online and large-scale settings.
\end{abstract}

\keywords{computing education, large online classes, data visualization projects}

\maketitle

\section{Introduction}

This paper outlines the key strategies, challenges faced, and solutions implemented in managing a 1000+ students (200+ teams of 4-6 students) in an online course, Data and Visual Analytics (CSE 6242) at Georgia Institute of Technology. Students work on projects to build, test, and deploy novel visualizations \cite{Scager16}. These projects are evaluated on both the novelty of algorithms and visualization techniques, and account for 50\% of the total course grade.\footnote{Different iterations of the CSE 6242 course, offered over the last 10+ years, can be accessed at \href{https://poloclub.github.io/\#cse6242}{https://poloclub.github.io/\#cse6242}}

\section{Process, Challenges and Key Strategies}
\subsection{Collaboration for Successful Team Formation}
The team formation process is student-driven~\cite{Scager16}, supported and facilitated by course staff through Ed Discussion\footnote{\url{https://edstem.org/us/}}. From the start of the course, students are encouraged to interact with one another to form project teams. The process is as follows: 

\begin{itemize}[topsep=2pt, itemsep=0mm, parsep=3pt, leftmargin=10pt]
    \item Introduction:  To create an inclusive environment, an introduction thread is created in Ed Discussion and all class participants are encouraged to introduce themselves and respond to others. TAs post their own introductions and respond to all student posts to foster engagement and emulate an in-person classroom. 
    \item Team Formation: To create a student-driven open communication channel, a team formation thread is created in Ed Discussion for students to share their subject area interests, team size needs, and time zone preferences. This allows teams to form based on shared interests and availability, promoting natural collaboration and engagement ~\cite{Nokes15}. Additionally, a Project Teams Spreadsheet tracks team formation. 
    Once teams are created, students register them in a spreadsheet, enabling course staff to track progress and gather team size and ungrouped student statistics, ensuring efficient placements.
    \item Ongoing Reminders: TAs actively monitor the process and send reminders to students, as needed. Additionally, TAs provide guidance on managing team dynamics, including how to structure early project meetings, team member roles and responsibilities, version control practices and guidelines for adjusting project scope based on data availability or any other issues.   
\end{itemize}
\subsection{Consistent Grading and Effective Feedback}
Teaching Assistants (TAs) are crucial to the success of the project implementation --- each semester a team of two professors and 25+ TAs support this process. They monitor team progress, provide grading and feedback on deliverables, resolve conflicts, and assist teams with technical or organizational challenges to ensure teams stay on track with project milestones. 
Project teams are required to submit deliverables at three key points of the project; these include a two-page proposal at the beginning, a four-page progress report at mid-point, and a six-page report, slides, as well as individual video presentations at the end of the project. 

\subsubsection{Constructive Feedback and Follow-Up}
One of the core foci of the grading process is to establish and maintain high-quality feedback provided by the TAs. Feedback is essential to guide students as they progress through their projects ~\cite{Burgess21}. Detailed feedback is given on each deliverable, with specific and actionable advice addressing both strengths and areas for improvement. TAs are instructed to avoid superficial comments like ``Good work'' and are expected to provide specific, actionable advice that addresses both strengths and areas for improvement. 

TAs are continuously reminded to keep feedback concise but meaningful, and they are encouraged to review their comments from earlier submissions to ensure continuity. If a team improves upon a previous submission, it is noted in the grading comments to acknowledge their growth and effort. 

\subsubsection{Consistent Grading}

A subset of TAs serve as Project Leads to monitor the grading process and to ensure consistent grading. Their responsibilities include: 

\begin{itemize}[topsep=2pt, itemsep=0mm, parsep=3pt, leftmargin=10pt]
    \item Monitoring team formation and course communication to ensure students receive notifications through multiple channels (Canvas\footnote{\url{https://canvas.instructure.com/}}, Ed Discussion, email). 
    \item Overseeing the grading process by ensuring consistent grading among TAs and that grading discrepancies are addressed early. Grading is completed independently by pairs of TAs, each assessing assigned group submissions without access to the other TA's grades. Once both TAs complete their evaluations, an assessment comparison is performed and any discrepancy over five points requires discussion to reconcile these differences. This process ensures consistency and fairness in the grading process ~\cite{Brinkley09}. 
    \item Normalizing the grades by reviewing all feedback and grades while ensuring consistency between the two. This process ensures that grades and feedback align with the course standards. 
    
\end{itemize}

\subsection{Facilitating 3000+ Peer Feedback}

The final report deliverables (slides and video presentation) are peer-reviewed by at least three randomly selected students from the course roster. This peer review promotes peer-to-peer learning, giving students a chance to engage with the work of their classmates and provide constructive feedback ~\cite{Burgess21}. Peer feedback is completed through the Peer Feedback module in the Canvas Learning Management System and includes: \textit{(i)} Peer grading: Students evaluate four presentations, each graded by at least three peers. \textit{(ii)} Quality Control: To ensure fairness, the final score is the average of the two highest peer scores, reducing outlier impact. \textit{(iii)} Constructive Feedback: Students provide feedback based on a rubric, focusing on clarity, project importance, \& visualizations.

After the peer grading period ends, the course staff reviews the grades and comments to identify any anomalies, such as large score variances or missing scores.  

The peer feedback system fosters peer-to-peer learning by exposing students to alternative approaches and a broader range of data visualization techniques and project ideas.

\subsection{Managing Team Dynamics and Conflicts}

Using our strategies to encourage collaboration, we have had great success in limiting the number of student conflicts. Most semesters, we see 0-3 groups with conflicts requiring course staff intervention. 
In such cases, the course staff step in to mediate. For example:

\begin{itemize}[topsep=2pt, itemsep=0mm, parsep=3pt, leftmargin=10pt]
    \item Communication to Teams: When a team reports unequal engagement, the teaching staff sends a message to all members, asking them to provide input and course corrections. This helps ensure that everyone is held accountable for their contributions. If necessary, individual emails are sent to specific team members to gather more detailed information about their involvement. 
    \item Team Conflict Resolution: When conflicts arise, course staff carefully review the situation, gathering input from all team members before making a final decision. The goal is to resolve issues in a way that ensures fairness while maintaining the integrity of the team’s work. If a team member is found to have contributed insufficiently, their grade may be adjusted accordingly.
    \item Internal Peer Feedback: Mid-semester, students complete an anonymous peer feedback survey to evaluate their teammates. This provides insights into team dynamics, helping course staff identify and address emerging issues early while encouraging teams to reflect on their collaboration and contributions.
\end{itemize}
Throughout this process, students learn valuable skills such as providing constructive feedback, collaborating in teams, and resolving conflicts --- skills that are highly sought-after in data science.

\section{Acknowledgments}
The authors thank the course staff for their vital contributions to the in-person and online course versions. This work is supported in part by the Google Award for Inclusion Research.

\bibliographystyle{ACM-Reference-Format}
\bibliography{references}


\begin{thebibliography}{4}


\ifx \showCODEN    \undefined \def \showCODEN     #1{\unskip}     \fi
\ifx \showDOI      \undefined \def \showDOI       #1{#1}\fi
\ifx \showISBNx    \undefined \def \showISBNx     #1{\unskip}     \fi
\ifx \showISBNxiii \undefined \def \showISBNxiii  #1{\unskip}     \fi
\ifx \showISSN     \undefined \def \showISSN      #1{\unskip}     \fi
\ifx \showLCCN     \undefined \def \showLCCN      #1{\unskip}     \fi
\ifx \shownote     \undefined \def \shownote      #1{#1}          \fi
\ifx \showarticletitle \undefined \def \showarticletitle #1{#1}   \fi
\ifx \showURL      \undefined \def \showURL       {\relax}        \fi
\providecommand\bibfield[2]{#2}
\providecommand\bibinfo[2]{#2}
\providecommand\natexlab[1]{#1}
\providecommand\showeprint[2][]{arXiv:#2}

\bibitem[Brindley et~al\mbox{.}(2015)]%
        {Brinkley09}
\bibfield{author}{\bibinfo{person}{J. Brindley}, \bibinfo{person}{L.~M. Blaschke}, {and} \bibinfo{person}{C. Walti}.} \bibinfo{year}{2015}\natexlab{}.
\newblock \showarticletitle{Creating Effective Collaborative Learning Groups in an Online Environment.}
\newblock \bibinfo{journal}{\emph{The International Review of Research in Open and Distributed Learning}} \bibinfo{volume}{10}, \bibinfo{number}{3} (\bibinfo{year}{2015}).
\newblock
\urldef\tempurl%
\url{https://eric.ed.gov/?id=EJ847776}
\showURL{%
\tempurl}


\bibitem[Burgess et~al\mbox{.}(2021)]%
        {Burgess21}
\bibfield{author}{\bibinfo{person}{Annette Burgess}, \bibinfo{person}{Chris Roberts}, \bibinfo{person}{Andrew~Stuart Lane}, \bibinfo{person}{Inam Haq}, \bibinfo{person}{Tyler Clark}, \bibinfo{person}{Eszter Kalman}, \bibinfo{person}{Nicole Pappalardo}, {and} \bibinfo{person}{Jane Bleasel}.} \bibinfo{year}{2021}\natexlab{}.
\newblock \showarticletitle{Peer review in team-based learning: influencing feedback literacy}.
\newblock \bibinfo{journal}{\emph{BMC Medical Educaton}} \bibinfo{volume}{21}, \bibinfo{number}{426} (\bibinfo{year}{2021}).
\newblock
\urldef\tempurl%
\url{https://doi.org/10.1186/s12909-021-02821-6}
\showDOI{\tempurl}


\bibitem[K et~al\mbox{.}(2016)]%
        {Scager16}
\bibfield{author}{\bibinfo{person}{Scager K}, \bibinfo{person}{Boonstra J}, \bibinfo{person}{Peeters T}, \bibinfo{person}{Vulperhorst J}, {and} \bibinfo{person}{Wiegant F.}} \bibinfo{year}{Winter 2016}\natexlab{}.
\newblock \showarticletitle{Collaborative Learning in Higher Education: Evoking Positive Interdependence.}
\newblock \bibinfo{journal}{\emph{CBE Life Sciences Education}} \bibinfo{volume}{14}, \bibinfo{number}{4} (\bibinfo{year}{Winter 2016}).
\newblock
\urldef\tempurl%
\url{https://doi.org/10.1187/cbe.16-07-0219}
\showDOI{\tempurl}


\bibitem[Nokes-Malach et~al\mbox{.}(2015)]%
        {Nokes15}
\bibfield{author}{\bibinfo{person}{T.J. Nokes-Malach}, \bibinfo{person}{J.E. Richey}, {and} \bibinfo{person}{S. Gadgil}.} \bibinfo{year}{2015}\natexlab{}.
\newblock \showarticletitle{When Is It Better to Learn Together? Insights from Research on Collaborative Learning.}
\newblock \bibinfo{journal}{\emph{Educational Psychology Review}}  \bibinfo{volume}{27} (\bibinfo{year}{2015}), \bibinfo{pages}{645--656}.
\newblock
\urldef\tempurl%
\url{https://doi.org/10.1007/s10648-015-9312-8}
\showDOI{\tempurl}


\end{thebibliography}

\end{document}